\documentclass[conference]{IEEEtran}
\IEEEoverridecommandlockouts
\usepackage{cite}
\usepackage{url}
\usepackage{amsmath,amssymb,amsfonts}
\usepackage{algorithm}
\usepackage{algorithmic}
\usepackage{graphicx}
\usepackage{subcaption}
\usepackage{textcomp}
\usepackage{xcolor}
\def\BibTeX{{\rm B\kern-.05em{\sc i\kern-.025em b}\kern-.08em
    T\kern-.1667em\lower.7ex\hbox{E}\kern-.125emX}}
\begin{document}

\title{Differences in Small-Signal Stability Boundaries Between Aggregated and Granular DFIG Models\\

\thanks{Correspondence: Dan Wu.}
}

\author{
\IEEEauthorblockN{Leyou Zhou$^1$, Mucheng Li$^1$, Xiaojie Shi$^1$, Meng Zhan$^1$, Juanjuan Wang$^2$, Dan Wu$^{1}$}
\IEEEauthorblockA{
\textit{1. School of Electrical and Electronic Engineering, Huazhong University of Science and Technology, Wuhan, China}\\
\textit{2. Department of Electrical Engineering, South China University of Technology, Guangzhou, China}
} 
\textit{\{leyou\_zhou, mucheng\_li, xiaojie\_shi, zhanmeng, danwuhust\}@hust.edu.cn and epjjwang@scut.edu.cn}
}

\maketitle

\begin{abstract}
Broadband oscillations in wind farms have been widely reported in recent years. 
Past studies have examined various types of oscillations in wind farms, relating small-signal stability to control settings, operating conditions, and electrical parameters. However, most analyses are performed on aggregated single-unit models, which may deviate from the true behavior, leading to misleading stability assessments. 
To investigate how aggregation affects stability conclusions, this paper develops detailed single-, two-, and three-unit doubly-fed induction generator (DFIG) models and their aggregated counterparts. Then, a D-decomposition-related ray-extrapolation method is proposed to characterize the small-signal stability region of nonlinear DFIG models in the parameter space, delineating stability boundaries under numerous parameter combinations.
The study reveals that aggregated models stability regions within the parameter planes of control settings and operating conditions differ from those of granular models in terms of basic shape, critical modes, and evolution patterns, posing a risk of misjudging stability margins.

\end{abstract}


\section{Introduction}
Oscillation events in wind farms have been widely reported in recent years with a broad spectrum of oscillatory frequencies ranging from a few Hz to thousands of Hz \cite{Oscilations}.
Various mechanisms of small-signal instability were investigated in wind farms under different network configurations. 

In the series-compensated configuration, sub-synchronous resonance (SSR) or sub-synchronous control interaction (SSCI) in DFIG-based wind farms are widely reported types of small-signal instability.
Based on an equivalent impedance model with the entire control loop, \cite{SeriesCompensated1} showed that the controller contributes to a negative resistance effect on system stability.
Moreover, the system's oscillation depends nonlinearly on operating conditions (e.g., wind speed, number of online turbines) and control settings (e.g., inner-loop proportional gains)\cite{SeriesCompensated2,SeriesCompensated5}.
Some studies showed that system damping varies non-monotonically with wind speed, and a higher series-compensation level can shrink the stable wind-speed range \cite{SeriesCompensated3}.
Other investigations revealed that wind farms with direct-drive permanent magnet synchronous generators (PMSG) in series-compensated networks can also be subject to SSR, whose dominant modes are strongly influenced by the grid-side control (GSC) settings \cite{SeriesCompensated4}. 


In the parallel-compensated configuration, the resonance risk in wind farms was explored in \cite{DFIG_ParallelCapacitor1, DFIG_ParallelCapacitor2,Type4_InsideMode2,ParallelCapacitor6}.
Based on impedance models, analysis revealed that the DFIG integrated weak power grid is particularly susceptible to high-frequency LC resonance under parallel compensated conditions  \cite{ParallelCapacitor6}. 
A weaker grid or a smaller compensated capacitor value can lead to a higher resonant frequency \cite{ParallelCapacitor6,DFIG_ParallelCapacitor2}. 
Moreover, LCL filters can significantly reshape the system’s high-frequency impedance compared to only L filters \cite{DFIG_ParallelCapacitor1,ParallelCapacitor6}. On the other hand, the proportional gain of the current control loop has a pronounced impact on the phase of the equivalent impedance \cite{DFIG_ParallelCapacitor1}, and digital control delay reduces the phase margin in the high-frequency band \cite{ParallelCapacitor6}. 
When considering PMSG-based wind farms connected to parallel-compensated networks, existing literature showed that wind speeds, DC-link capacitance, GSC DC-voltage PI gains, grid-side filter inductance, and PLL settings are responsible for the shift of participation factors among different modes \cite{Type4_InsideMode2}. 

Most prior studies adopted aggregated wind farm models (typically capacity-weighted aggregation), which can reduce a multi-unit wind farm to a single unit \cite{kunjumuhammed2016adequacy}, largely reducing analytical difficulties. 
However, whether such a simplification can fully replicate all the risks of small-signal instability remains undecided. 
It is noted that weakly or negatively damped oscillatory modes can exist among multiple wind turbines in a wind farm \cite{Type4_InsideMode2, DFIG_InsideMode1, Type4_InsideMode1}. 
The modal analysis further showed that the damping of the aggregated model can be altered and spurious composite modes can be introduced, leading to unreliable stability assessment  \cite{SeriesCompensated3, kunjumuhammed2016adequacy}. Those findings between aggregated models and their granular counterparts were mainly based on a preselected operating condition and control setting, leaving the global properties of the entire small-signal stability region in multi-dimensional parameter space unexplored.

Motivated by the above gap, this article aims to justify the validity of the aggregated model from a global perspective. It first builds detailed nonlinear models for single-, two-, and three-unit DFIG systems and their aggregated counterparts. Then, an extended D-decomposition method \cite{D-com} is utilized to mathematically characterize the stability boundary in the parameter space of DFIG systems. A ray extrapolation algorithm with boundary correction (REBC) is further proposed to reveal the geometric properties of the stability boundary for diverse parameter combinations. 
This study offers new insights into the validity of small-signal stability assessment for DFIG wind farm models.


\section{Dynamic Model of DFIG System}
\label{sec:model}
\subsection{Wind System Configuration}
\begin{figure}[tb]
    \centering
    \includegraphics[width=1\linewidth]{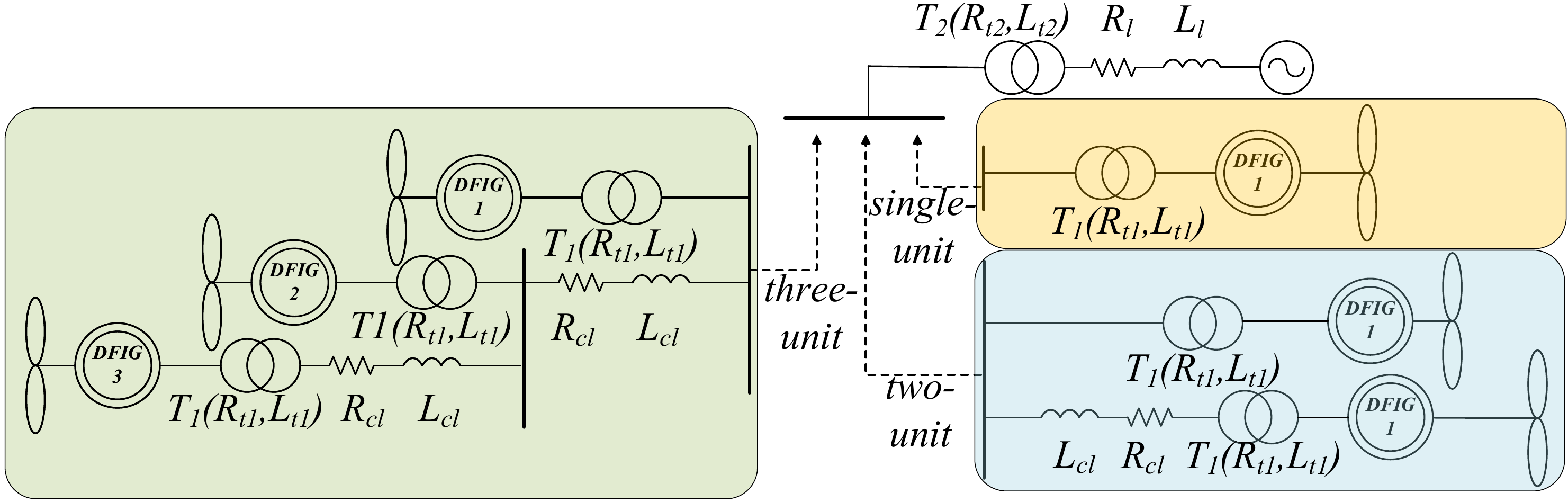}
    \caption{Schematic diagram of DFIGs systems}
    \label{fig:SystemTopology}
\end{figure}

We consider the tree structure network which is shown in Fig.\ref{fig:SystemTopology}. 
The turbine circuit and control diagrams of the DFIG unit are shown in Fig.\ref{fig:DFIG_Control_block}. 
Each unit comprises a wind turbine, drive-train, induction generator, rotor-side converter (RSC), grid-side converter (GSC), AC filter, and the associated control system, with parameters mainly taken from\cite{DFIG_parameter}. 
Throughout this paper, we denote subscripts $s$, $r$, and $g$ as stator, rotor, and grid side quantities, respectively. 
Subscripts $d$ and $q$ as the d-axis and q-axis components in the $dq$ reference frame. 
In the synchronous rotating $xy$ framework, the $x$ axis is aligned with the grid voltage phasor $\boldsymbol U_{g}$. The controllers are designed in the $dq$ frame from a PLL tracking the PCC voltage angle.
Both the RSC and GSC employ conventional vector control: the RSC regulates generator speed and reactive power, whereas the GSC regulates the DC-link capacitor voltage and provides reactive power support at the PCC.

\subsection{Dynamic Model of DFIG}

\paragraph{Induction Machine}
Under the motor sign convention 
\begin{subequations}
\begin{align}
u_{sy}&=R_{s}i_{sy}+\omega_b^{-1}p\psi_{sy}+\omega_{s}\psi_{sx} \\
u_{sx}&=R_{s}i_{sx}+\omega_b^{-1}p\psi_{sx}-\omega_{s}\psi_{sy} \\
u_{ry}&=R_{r}i_{ry}+\omega_b^{-1}p\psi_{ry}+(\omega_{s}-\omega_{r})\psi_{rx} \\
u_{rx}&=R_{r}i_{rx}+\omega_b^{-1}p\psi_{rx}-(\omega_{s}-\omega_{r})\psi_{ry} \\
\psi_{sy}&=L_{s}i_{sy}+L_{m}i_{ry} \\
\psi_{sx}&=L_{s}i_{sx}+L_{m}i_{rx} \\
\psi_{ry}&=L_{m}i_{sy}+L_{r}i_{ry}\\
\psi_{rx}&=L_{m}i_{sx}+L_{r}i_{rx}
 \end{align}
\end{subequations}
where $p=\frac{d}{dt}$;
 $\omega_b$ is the nominal (synchronous) electrical angular frequency ($\omega_b=100\pi\ \text{rad/s}$ for a 50-Hz system);
 $\omega_s$ and $\omega_r$ are the stator and rotor electrical angular frequencies; $u_{sx},u_{sy},u_{rx},u_{ry}$ are the $x$ and $y$ axis components of the stator/rotor voltage; 
 $i_{sx},i_{sy},i_{rx},i_{ry}$ are the $x$ and $y$ axis components of the stator/rotor currents; 
$\psi_{sx},\psi_{sy},\psi_{rx},\psi_{ry}$ are the $x$ and $y$ axis components of the stator/rotor flux linkage; 
 $R_s$ and $R_r$ are the  stator and rotor resistance;
 $L_s$ and $L_r$ are the  stator and rotor inductance;
 and $L_m$ is the magnetizing inductance.

\begin{figure}[tb]
    \centering    \includegraphics[width=1\linewidth]{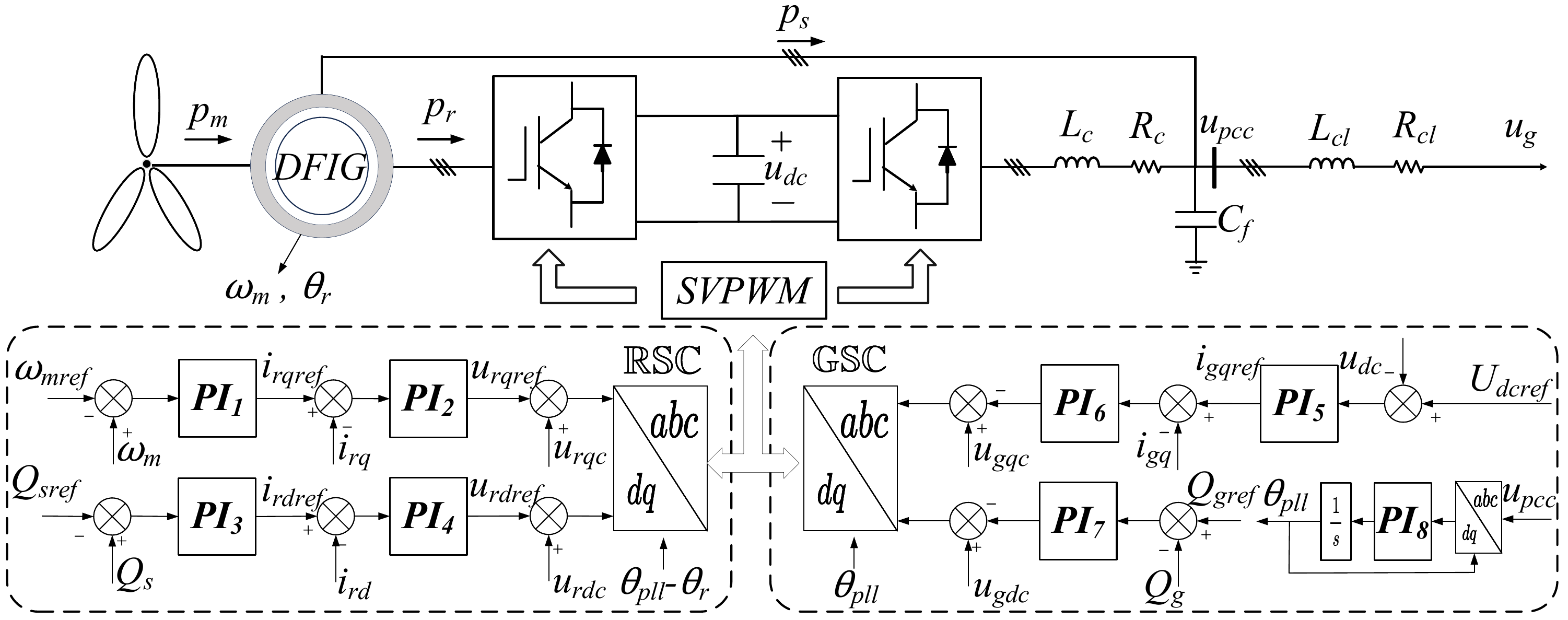}
    \caption{Control configuration for a single-unit DFIG}
\label{fig:DFIG_Control_block}
\end{figure}

The DFIG drive-train is modeled as a single-mass. 
\begin{equation}
p\omega_{\mathrm{r}}=\frac{T_{\mathrm{e}}-P_{\mathrm{m}}/{\omega_{\mathrm{r}}}}{2H}
\end{equation}%
where $T_e$ denotes the electromagnetic torque, $P_m$ the mechanical input power and $H$ the inertia constant.

\paragraph{Grid Side Converter and Filter}
The main circuit model of GSC, together with the terminal (AC-side) filter capacitor, is formulated below. 
\begin{subequations}
\begin{align}
{L_{c}}\omega_b^{-1}pi_{gx}&=u_{sx}-u_{gx}-R_{c}i_{gx}+\omega_{s}L_{c}i_{gy}\\
{L_{c}}\omega_b^{-1}pi_{gy}&=u_{sy}-u_{gy}-R_{c}i_{gy}-\omega_{s}L_{c}i_{gx}\\
{C_{dc}u_{dc}} pu_{dc}&={P_{in}-P_{out}}\\
{C_{f}}\omega_b^{-1}{pu_{sx}}&=i_{lx}-i_{sx}-i_{gx}+C_{f}\omega_{s}u_{sy}\\
{C_{f}}\omega_b^{-1}{pu_{sy}}&=i_{ly}-i_{sy}-i_{gy}+C_{f}\omega_{s}u_{sx}
\end{align}
\end{subequations}
where $u_{sx}$ and $u_{sy}$ represent the stator voltage $xy$ components; 
$u_{gx}$ and $u_{gy}$ denote the GSC average voltage generated by SVPWM; 
$u_\mathrm{dc}$ is the DC-link capacitor voltage; 
$i_\mathrm{gx}$ and $i_\mathrm{gy}$ are the GSC current $xy$ components;  $i_{lx}$ and $i_{ly}$ are the cable-line current $xy$ components.
$R_c$, $L_c$, and $C_f$ denote the filter resistance, inductance, and capacitance, respectively.

\paragraph{Transmission line}
A generic AC line model is adopted to provide a unified representation:
\begin{subequations}
\begin{align}
{L}\omega_b^{-1}pi_{lx}&=u_{lx1}-u_{lx2}-Ri_{lx}+\omega_{s}Li_{ly}\\
{L}\omega_b^{-1}pi_{ly}&=u_{ly1}-u_{ly2}-Ri_{ly}-\omega_{s}Li_{lx}
\end{align}
\end{subequations}
where $u_{lx1}$, $u_{ly1}$,  $u_{lx2}$, and $u_{ly2}$ denote the terminal voltages of the line in $xy$ framework; $R$ and $L$ are resistance and inductance of the line.

\paragraph{GSC and RSC Control} 
$K_{pn}$ and $K_{in} ~(n = 1, 2, \ldots, 8)$ denote the proportional and integral gains of the n-th PI controller in Fig.\ref{fig:DFIG_Control_block}, respectively;
 $x_n (n = 1, 2, \ldots, 8)$ denotes the corresponding integrator state. 

On the RSC, the $q$-axis loop forces the rotor speed to follow its reference:
\begin{subequations}
 \begin{align}
    px_1&=K_{i1}(\omega_{m}-\omega_{mref})\\
    px_2&=K_{i2}(i_{rqref}-i_{rq})\\
    i_{rqref}&=x_{1}+K_{p1}(\omega_{m}-\omega_{mref})\\
    u_{rqref}&=x_{2}+K_{p2}(i_{rqref}-i_{rq})+u_{rqc}\\
       u_{rqc}&=\frac{L_m}{L_s} \left( u_{sq} - \omega_r \psi_{sd} \right)+ \omega_{sl} \left( L_r - \frac{L_m^2}{L_s} \right) i_{rd}
    \end{align}
\end{subequations}
where $\omega_{sl}=\omega_s-\omega_r$.
The $d$-axis loop regulates stator reactive power to its reference:
\begin{subequations}
 \begin{align}
    px_3&=K_{i3}(Q_{s}-Q_{sref})\\
    px_4&=K_{i4}(i_{rdref}-i_{rd})\\
   i_{rdref}&=x_{3}+K_{p3}(Q_{s}-Q_{sref})\\
    u_{rdref}&=x_{4}+K_{p4}(i_{rdref}-i_{rd})+u_{rdc}\\
    u_{rdc}&=\frac{L_m}{L_s}\omega_{sl}  \psi_{sq} - \omega_{sl}\left( L_r - \frac{L_m^2}{L_s} \right) i_{rq}
    \end{align}
\end{subequations}
On the GSC, the $q$-axis loop regulates the capacitor voltage:
\begin{subequations}
 \begin{align}
    px_5&=K_{i5}(u_{dcref}-u_{dc})\\
    px_6&=K_{i6}(i_{gqref}-i_{gq})\\
    i_{gqref}&=x_{5}+K_{p5}(u_{dcref}-u_{dc})\\
    u_{gqref}&=-x_{6}-K_{p6}(i_{gqref}-i_{gq})+u_{gqc}\\
    u_{gqc}&=u_{sq}-\omega_{s}L_{c}i_{gd}
    \end{align}
\end{subequations}
The $d$-axis loop regulates reactive power:
\begin{subequations}
 \begin{align}
    px_7&=K_{i7}(Q_{gref}-Q_{g})\\
    u_{gdref}&=-x_{7}-K_{p7}(Q_{gref}-Q_{g})+u_{gdc}\\
    u_{gdc}&=u_{sd}+\omega_{s}L_{c}i_{gq}
    \end{align}
\end{subequations}



\paragraph{Phase-lock loop}
The dynamic PLL model is 
\begin{subequations}
    \begin{align}
            p\delta&=x_{8}-K_{p8}(u_{sx}\cos\delta+u_{sy}\sin\delta)\\
px_{8}&=-K_{i8}(u_{sx}\cos\delta+u_{sy}\sin\delta)\\
\delta&=\theta_{pll}-\theta_{xy}
    \end{align}    \label{eq:PLL}%
\end{subequations}%
where $\delta$ denotes the phase-lead angle of the PLL-generated $dq$ reference frame relative to the $xy$ reference frame aligned with grid-voltage phasor $\boldsymbol{U_g}$.

\paragraph{Unified state-space formulation of the system}
As detailed above, the AC transmission network and the electromechanical dynamics of each DFIG are formulated in a common rotating $xy$ reference frame, whereas the DFIG control subsystems are represented in their respective $dq$ reference frames.
To match different frameworks, the port-voltage and port-current components of each element must be transformed between their device-specific $dq$ frames and the common $xy$ frame. 
\begin{equation}
\label{eq:transform}
    \begin{pmatrix}
    A_x\\
    A_y
    \end{pmatrix} = \begin{pmatrix}
    \cos\delta & -\sin\delta \\
    \sin\delta & \cos\delta
    \end{pmatrix} \begin{pmatrix}
    A_d\\ 
    A_q
    \end{pmatrix}
\end{equation}
where $A$ denotes port-quantity.
Accordingly, the dynamic models can be formulated in the ODE form:
\begin{equation}
    \dot{x}=f(x,t)
    \label{eq:ode}
\end{equation}
Furthermore, the aggregated model is obtained by applying the method in \cite{ kunjumuhammed2016adequacy}.
The aggregation includes the DFIG, the interfacing transformer, and the collector network.
Assuming all units are the same, the per-unit parameters of the aggregated DFIG are kept identical to those of a single unit\cite{ kunjumuhammed2016adequacy}. The interfacing transformer and collector network parameters are scaled to preserve power losses between the detailed and aggregated model.

\subsection{Model Validation}
\begin{figure}[tb]
\centering
\begin{subfigure}[t]{0.48\linewidth}
    \includegraphics[width=\linewidth]{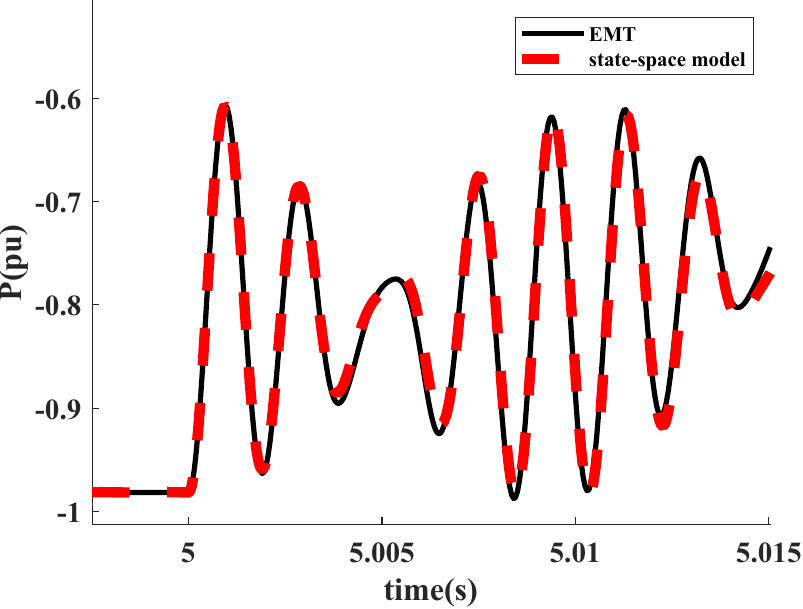}
    \caption{Time-domain response under a voltage dip to 0.8 pu}\label{fig:time-domain waveforms}
\end{subfigure}\hfill
\begin{subfigure}[t]{0.48\linewidth}
    \includegraphics[width=\linewidth]{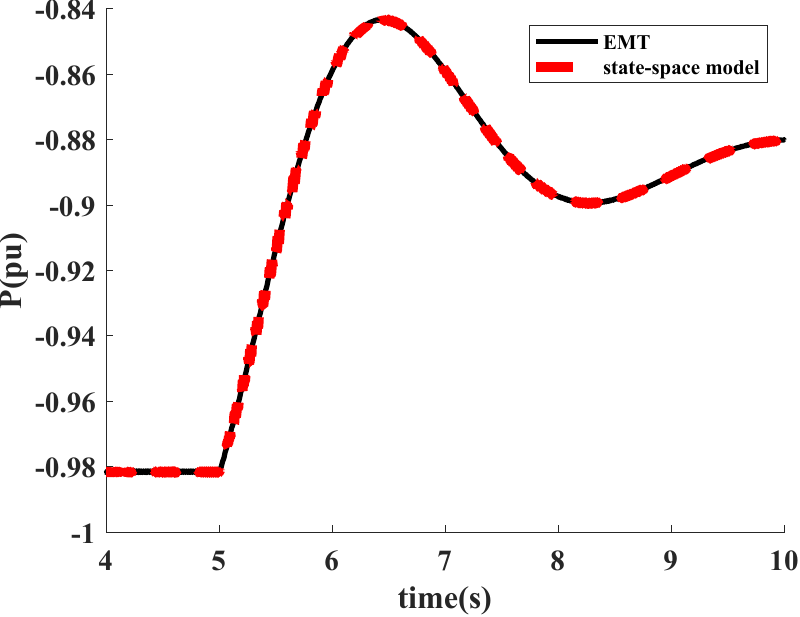}
    \caption{Time-domain response under a mechanical-power dip to 0.9 pu}\label{fig:spectral analysis}
\end{subfigure}
\caption{Validation of EMT and state-space model responses. }
\label{fig:validation}
\end{figure}

To validate the state-space model, an electromagnetic-transient (EMT) simulation benchmark is implemented in Matlab/Simulink. 
Taking the single-unit case as an example, a voltage drop to 0.8 pu and a mechanical power drop to 0.9 pu are applied at $t=5\ \text{s}$, respectively. 
In Fig.\ref{fig:validation}, the time-domain simulation results confirm the fidelity of the above model.

\section{Constructing Small-Signal Stability Region}
\label{method}
Given a system depending on a parameter vector $k\in \mathbb{R}^m$, the D-decomposition method analyzes
the characteristic polynomial $P(j\omega,k)$, with the stability region in the parameter space denoted by (\ref{eq:D-dcomposition}).\begin{equation}\label{eq:D-dcomposition}
    P(j\omega,k)=0,-\infty<\omega<+\infty
\end{equation}
This approach essentially maps the imaginary axis (the boundary of instability) into the parameter space.
The curve $k(\omega)$ partitions the parameter space into distinct regions, with each containing a fixed number of stable and unstable eigenvalues.


For high-dimensional nonlinear state-space models(DFIGs), the explicit expansion of the characteristic polynomial yields prohibitively complex algebraic expressions.
Furthermore, the intricate coupling between the coefficients and the system states, control parameters, and network parameters is difficult to characterize, posing significant challenges for both theoretical derivation and numerical implementation.
To address these issues, additional real eigenvalue parameter dimensions  are introduced to construct a state-parameter-eigenvalue Cartesian product space$(x,k,u,v,\omega)\in \mathbb{R}^{n\times m\times n\times n\times1}$. The equilibrium manifold of the system defined in  (\ref{eq:ode}) is then embedded into this augmented space. The following conditions are examined to unify various bifurcation criteria, thereby establishing a high-dimensional yet low-order mathematical model for the stability boundary.
\begin{subequations}
    \begin{align}
f_x(x,k)u+\omega v=0\\
f_x(x,k)v-\omega u=0\\
u^Tu+v^Tv-1=0
    \end{align}    \label{eq:boundarycondition}%
\end{subequations}%
where $f_x(x,k)$ specifies the jacobian matrix $\partial f/\partial x$; $\omega$ represents the unstable frequency; $v$ and $u$ denote the real and imaginary parts of the eigenvector.
Subsequently, a set of parameter direction constraints is introduced to characterize the specific trajectory of system parameters variations, 
\begin{equation}
k - k_0 - s\mathbf{d}(\theta)=0
\label{eq:direction}
\end{equation}
where $k$ is the parameter vector with its initial value at $k_0$; $s$ is the Euclidean distance in the parameter space; 
$\mathbf{d}(\theta)$ is the direction vector determined by the angle parameter $\theta$ ($\mathbf{d}(\theta)=[\cos\theta,\sin\theta]^T$ in two parameter plane).
By integrating (\ref{eq:ode}) (\ref{eq:boundarycondition}) and (\ref{eq:direction}), the complete mathematical model for characterizing the system stability boundary is formulated (\ref{eq:boundary model}).

\begin{equation}\label{eq:boundary model}
F(z)=\mathbf{0},
z=(x,v,u,k,s,\omega,\theta)
\end{equation}%
 
By sweeping the parameter space, the stability region is characterized. 
A predictor–corrector approach is adopted. The predictor advances along rays and the corrector forces the boundary-equation solving (Algorithm \ref{algorithm:Ray-based}), yielding the stability region in the parameter space. 
To ensure that each parameter $\tilde{k}$ (in p.u.) contributes at most
unity over its admissible range to the weighted search distance, we specify the original range $(\tilde{k}_{lb},\tilde{k}_{ub})$ and normalize by $\Delta=\tilde{k}_{ub}-\tilde{k}_{lb}$, i.e., $k=\tilde{k}/\Delta$ with $(k_{lb},k_{ub})=(\tilde{k}_{lb}/\Delta,\tilde{k}_{ub}/\Delta)$, so that $k_{ub}-k_{lb}=1$.

\begin{algorithm}
\caption{Constructing Ray Extrapolation Region}\label{algorithm:Ray-based}
    \renewcommand{\algorithmicrequire}{\textbf{Input:}}
    \renewcommand{\algorithmicensure}{\textbf{Output:}}
\begin{algorithmic}[1]
\REQUIRE Initial operating point $x_{0}\in R^{n}$, initial parameter $k_{0}\in R^{m}$, parameter bound $(k_{lb},k_{ub})$, initial distance  $s_{0}=0.1$, step-size growth rate $\alpha=1.05$.
\ENSURE Distance set to stability-region boundary $S^{*}$, unstable frequency set $\Omega^{*}$
\FOR{$i = 1,2\dots100$, $\theta_{i}=i\cdot\pi/50$}
\STATE $x=x_{0},s=s_{0}$
\WHILE{true}
\STATE $s\gets \alpha s,k\gets k_{0}+ s\mathbf{d}(\theta_i) $
\STATE Stability Check:$\lambda$ is eigenvalue
\STATE $(x,\lambda,\omega,v,u) \gets \textsc {Solve} \bigl(f(x,k)=0\bigl)$
\STATE $k^{+}=(k-k_{ub})_{+},k^{-}=(k_{lb}-k)_{+}$
\IF{$\max(\left\| k^{+} \right\|_\infty,\left\| k^{-} \right\|_\infty)>0\textbf{ and }\max Re(\lambda)<0$}
\STATE $S^{*}(i)\gets Smax(i)$, $\Omega^*(i)\gets NaN$
\STATE \textbf{break}   
\ELSIF{$\max Re(\lambda)>0$}

\STATE $(S^{*}(i),\Omega^*(i)) \gets {Solve} \bigl(F(x,v,u,k,s,\omega,\theta_{i})=0\bigl)$
\STATE \textbf{break}   
\ENDIF
\ENDWHILE
\ENDFOR
\RETURN Outputs
\end{algorithmic}
\end{algorithm}

\section{Numerical Results}
\label{case study}
In this section, we consider the entire parameter space, including operation conditions, control settings, and electrical parameters for the detailed single-, two-, and three-unit models and their aggregated counterparts. 
The interconnection topology among the detailed or aggregated units and the grid maintains across all cases (Fig.\ref{fig:SystemTopology}).
In addition, all parameters are normalized, and each parameter’s original admissible range $(\tilde{k}_{lb},\tilde{k}_{ub})$ is given in the figure caption. Certain characteristics on the small-signal stability boundary are observed and visualized from projections on 2-dimensional parameter slices. 
Specifically, a few typical 2D instances are shown below. We use hexagram marks to show the original parameter setting, whereas black pentagrams indicate the prescribed parameter bounds. 
The small-signal stability regions are enclosed by colorful curves, whose colors correspond to the heatmap of the unstable frequency spectrum.

Across a broad range of parameter combinations, the stability boundaries exhibit recurring features as the model changes: boundary morphological evolution and unstable mode switching, as demonstrated below. 

\subsection{Morphological features of the stability boundary}

\begin{figure}[tb]
\centering
\begin{subfigure}[t]{0.49\linewidth}
     \includegraphics[width=\linewidth]{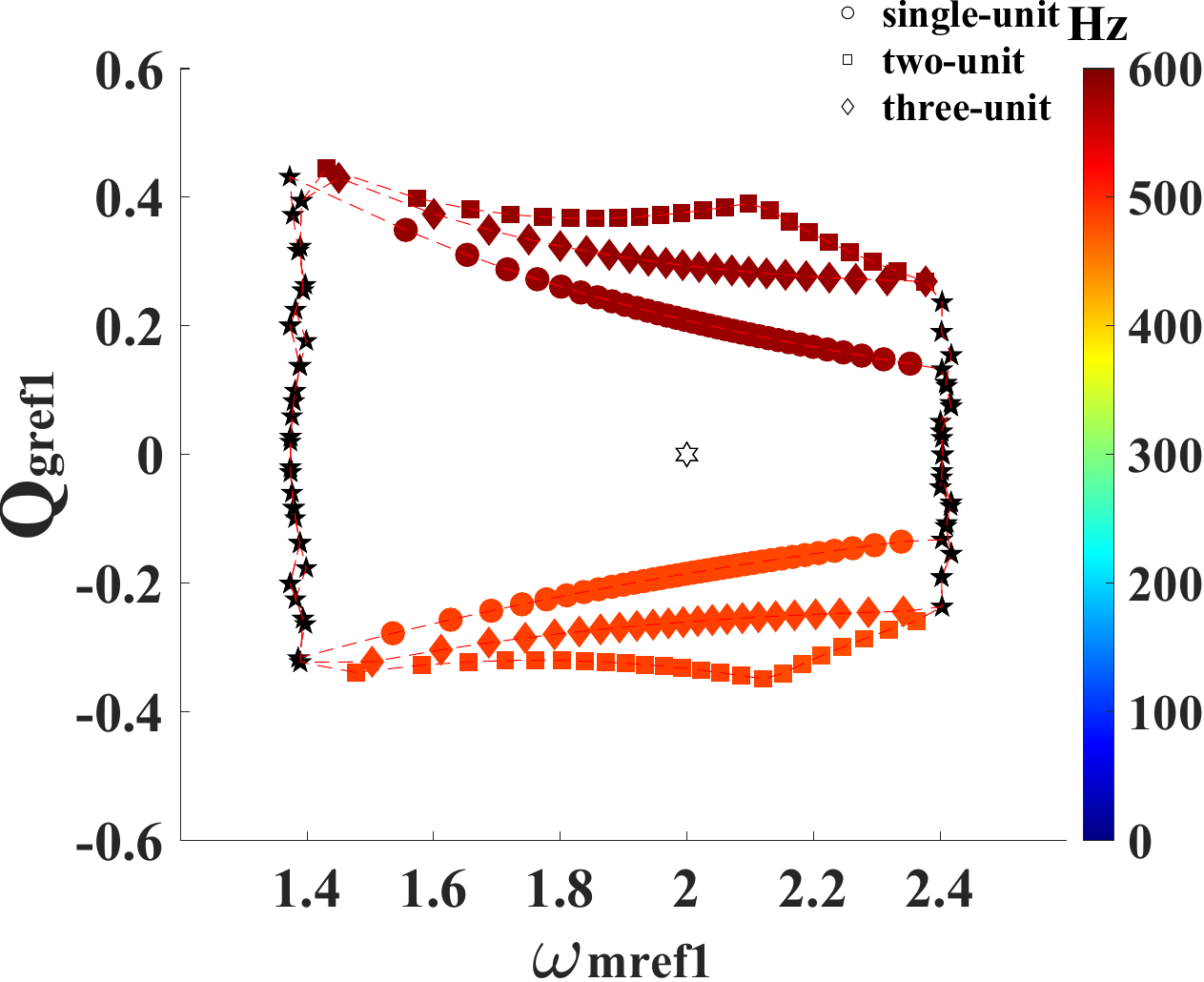}
    \caption{Stability boundary of granular models}
    \label{fig:a1}
\end{subfigure}\hfill
\begin{subfigure}[t]{0.49\linewidth}
    \includegraphics[width=\linewidth]{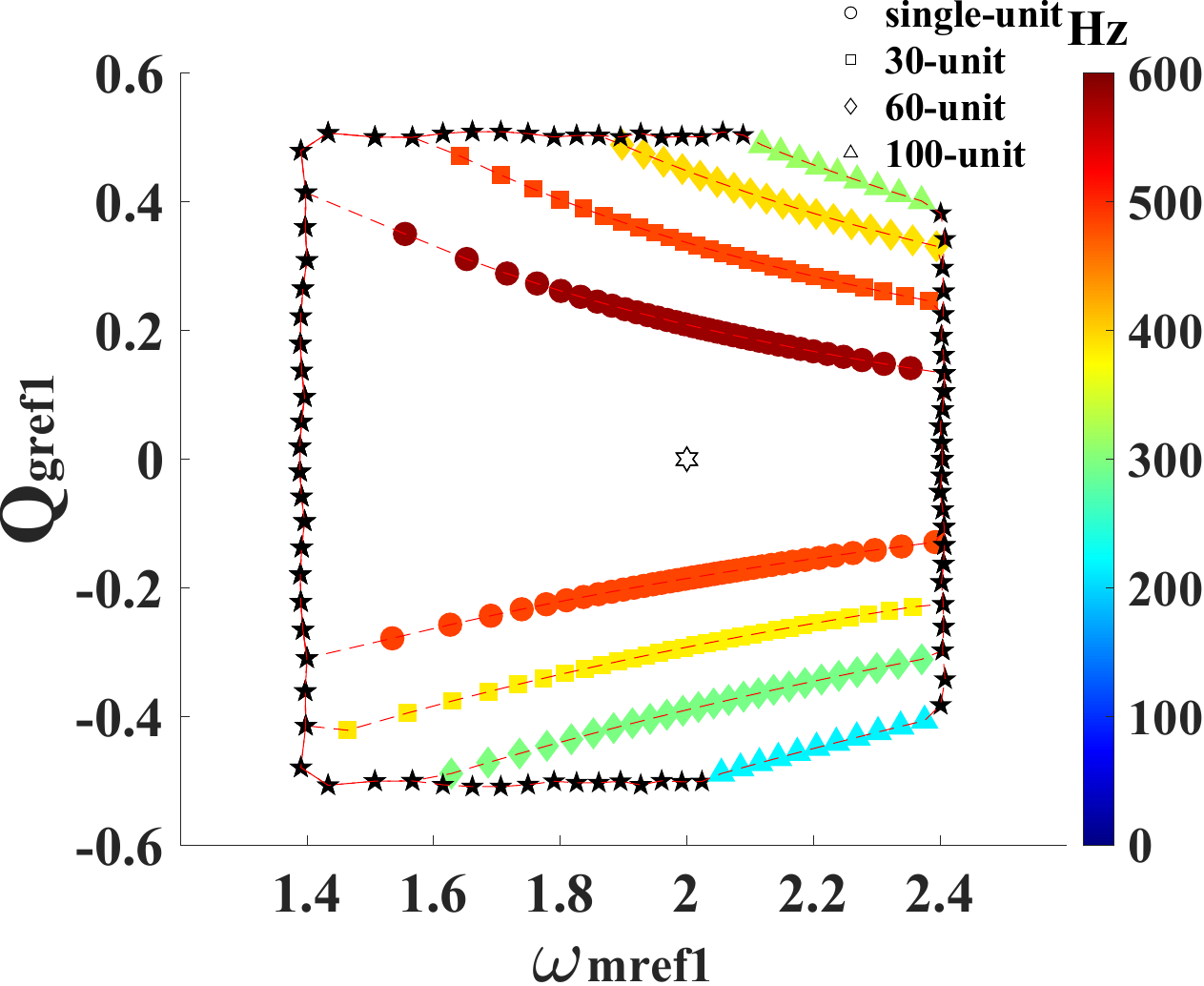}
    \caption{Stability boundary of single-unit aggregated models}\label{fig:a2}
\end{subfigure}
\begin{subfigure}[t]{0.49\linewidth}
     \includegraphics[width=\linewidth]{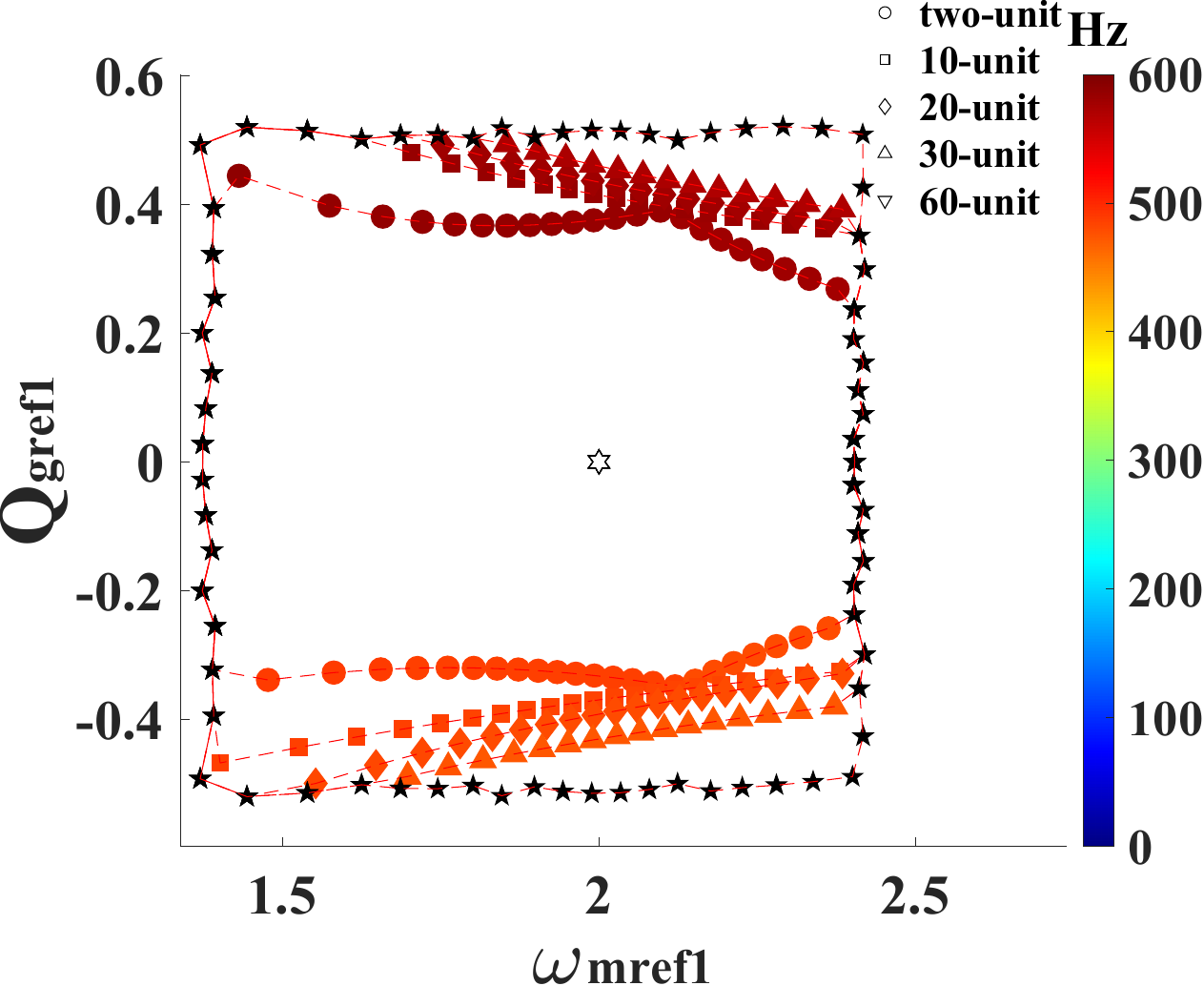}
    \caption{Stability boundary of two-unit aggregated models}\label{fig:a1}
\end{subfigure}\hfill
\begin{subfigure}[t]{0.49\linewidth}
    \includegraphics[width=\linewidth]{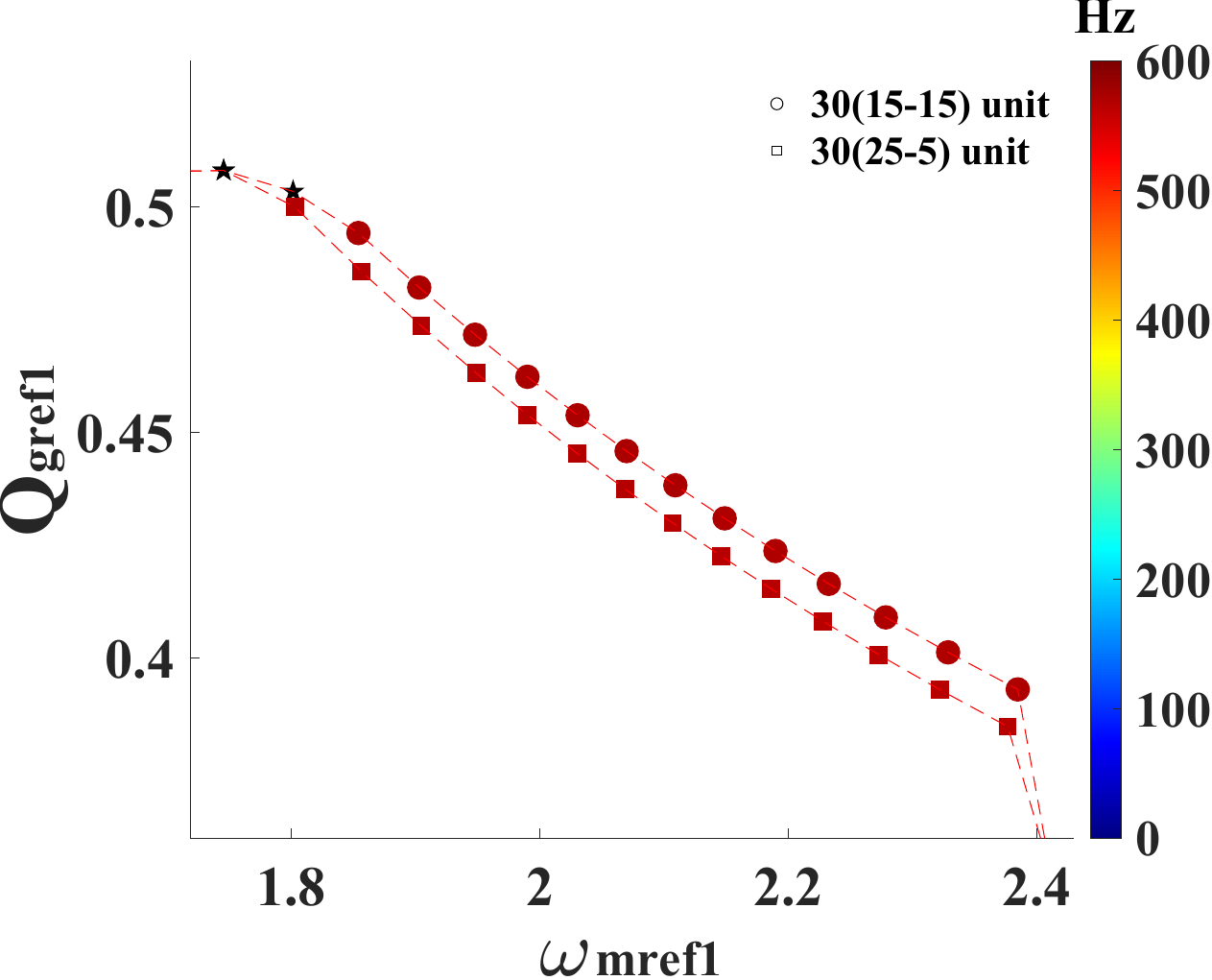}
    \caption{Effect of allocation ratio on the stability boundary}
    \label{fig:a2}
\end{subfigure}
\caption{Morphological evolution of the stability boundary in $\omega_{mref}-Q_{gref}$ (normalized parameter plane), the original parameter ranges are
$\tilde{\omega}_{mref}\text{(p.u.)}\in [0.7 ,1.2]$ and $\tilde{Q}_{gref} \text{(p.u.)} \in [-0.2,+0.2]$
}
\label{fig:MorphologicalEvolution}
\end{figure}


Compared to the single-unit system, the two-unit system shows a marked expansion of the stability region: a pronounced cusp appears at $(2.09,0.39)$ in Fig.\ref{fig:MorphologicalEvolution} (a). 
However, the stability region of the three-unit system lies between that of the single- and two-unit cases, suggesting that \textit{the small-signal stability region does not vary monotonically as the number of granular units increases.} 

When we continuously aggregate more units into a single-machine model, the stability region, shown in Fig.\ref{fig:MorphologicalEvolution} (b), merely expands outward, failing to capture the boundary-shape variations seen in the detailed models in Fig.\ref{fig:MorphologicalEvolution} (a). It suggests that \textit{the granular model can admit more dynamic geometries on the stability boundary than the simply aggregated single-unit model, casting doubt on the validity of the linearized boundaries obtained from the aggregated model.} 

On the other hand, the aggregated single-unit model exhibits a profound unstable mode frequency drift as more units are aggregated together, shown in Fig.\ref{fig:MorphologicalEvolution} (b). However, as we aggregate more units into the two-unit model with an even allocation ratio in each aggregated representation, shown in Fig.\ref{fig:MorphologicalEvolution} (c), the aggregated system does not present a substantial unstable mode frequency drift. It indicates that \textit{different basic configurations for aggregation can result in very different unstable mode frequencies.} Another interesting finding in Fig.\ref{fig:MorphologicalEvolution} (d) shows that whether we evenly aggregate units to the two-unit model or make it uneven only changes the stability boundary mildly, suggesting that \textit{the basic system configuration is more influential than how to allocate units for aggregation.} 


\subsection{Critical mode switching on the stability boundary}
\begin{figure}[tb]
\centering
\begin{subfigure}[t]{0.49\linewidth}
     \includegraphics[width=\linewidth]{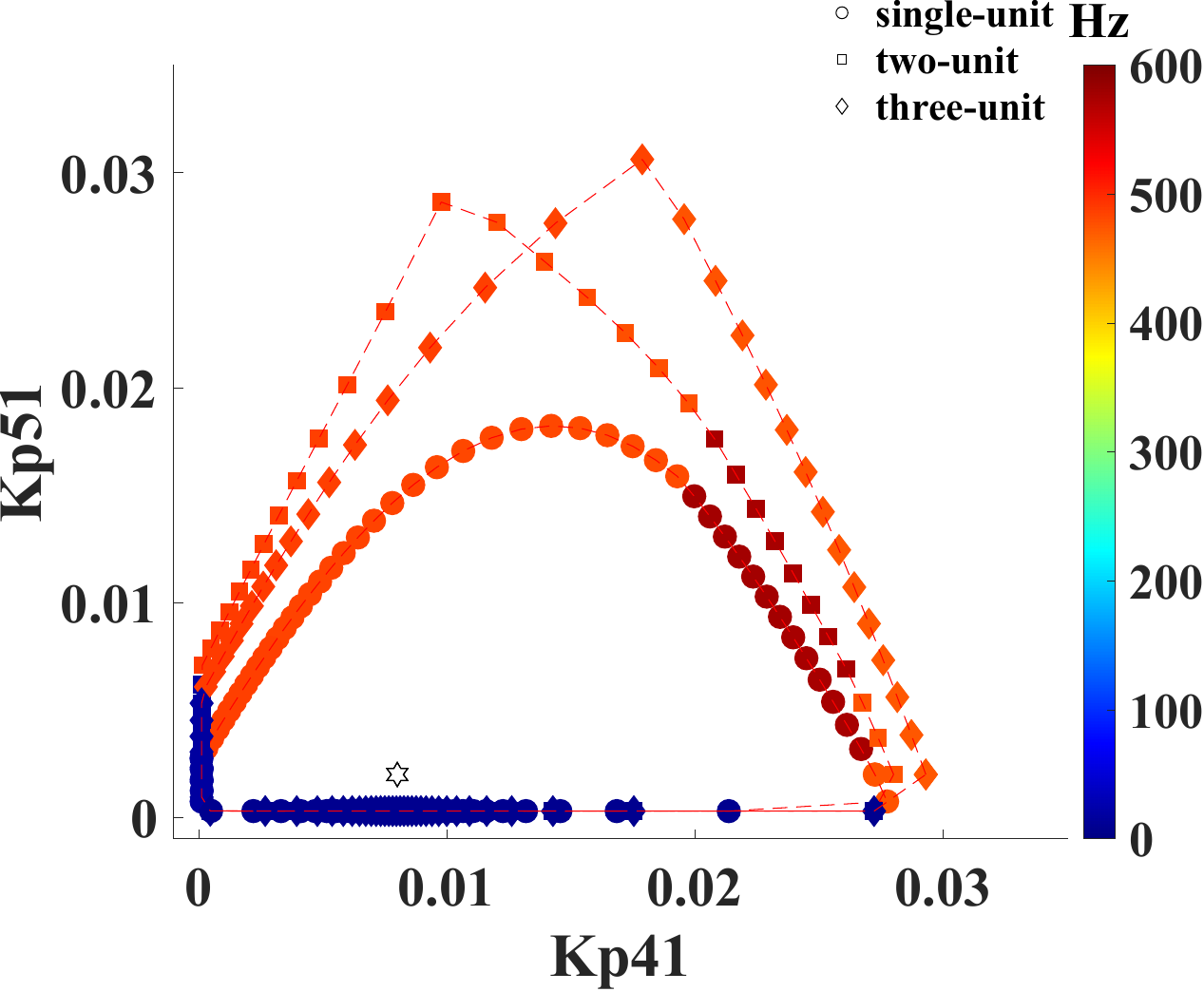}
    \caption{Stability boundary of granular models}\label{fig:a}
\end{subfigure}\hfill
\begin{subfigure}[t]{0.49\linewidth}
    \includegraphics[width=\linewidth]{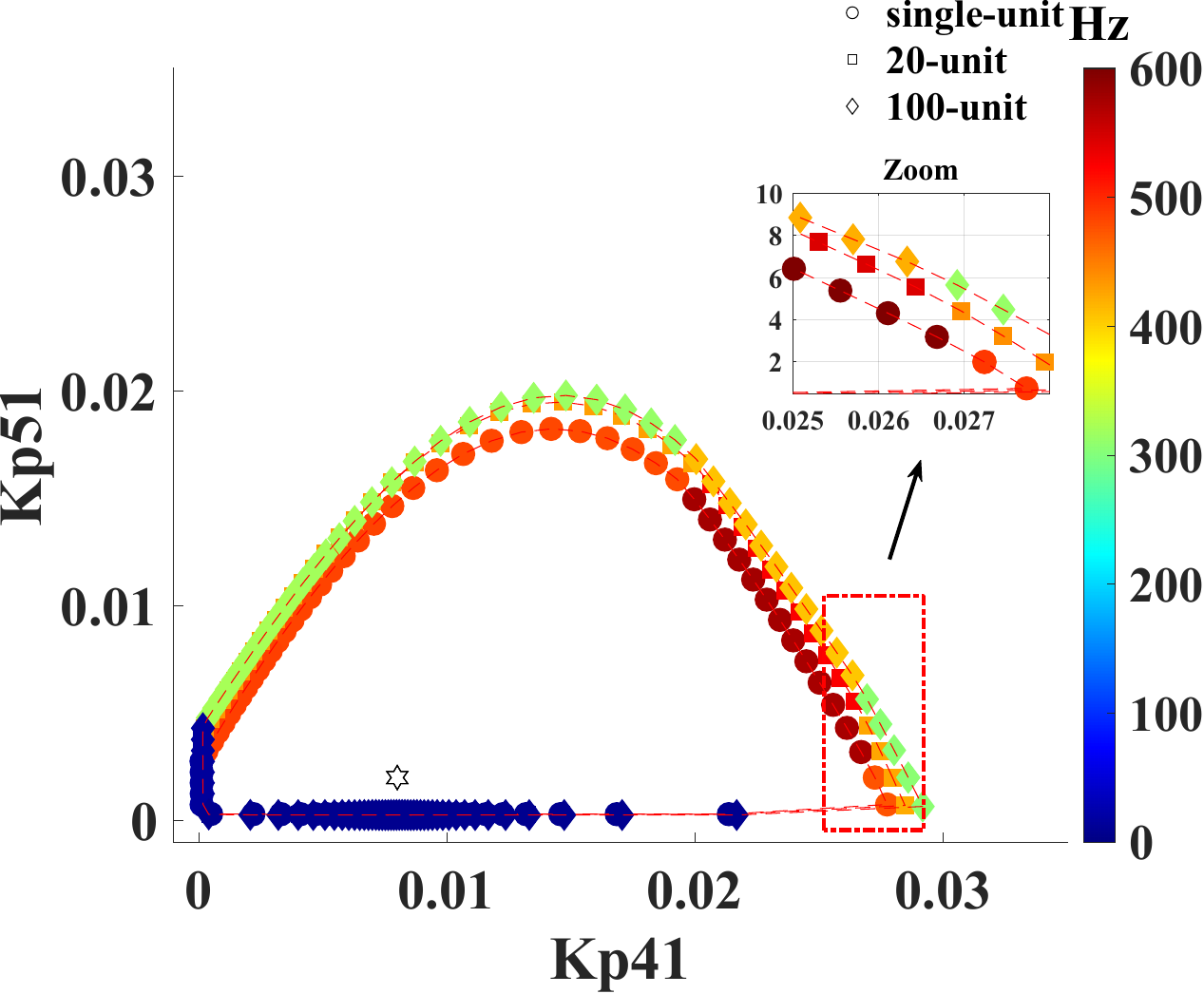}
    \caption{Stability boundary of single-unit aggregated models}\label{fig:b}
\end{subfigure}
\begin{subfigure}[t]{0.49\linewidth}
     \includegraphics[width=\linewidth]{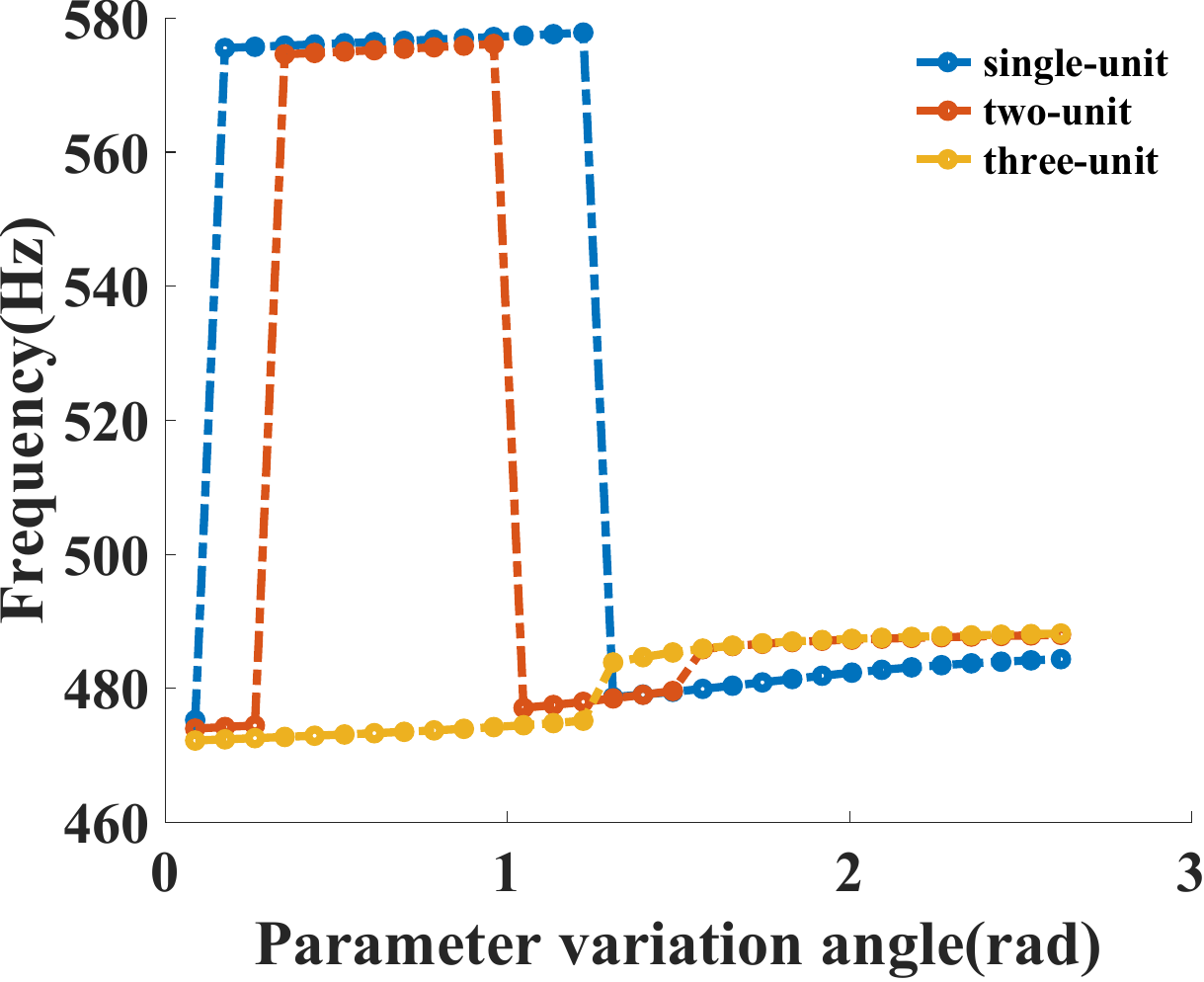}
    \caption{Critical modes switching}
    \label{fig:c}
\end{subfigure}\hfill
\begin{subfigure}[t]{0.49\linewidth}
\includegraphics[width=\linewidth]{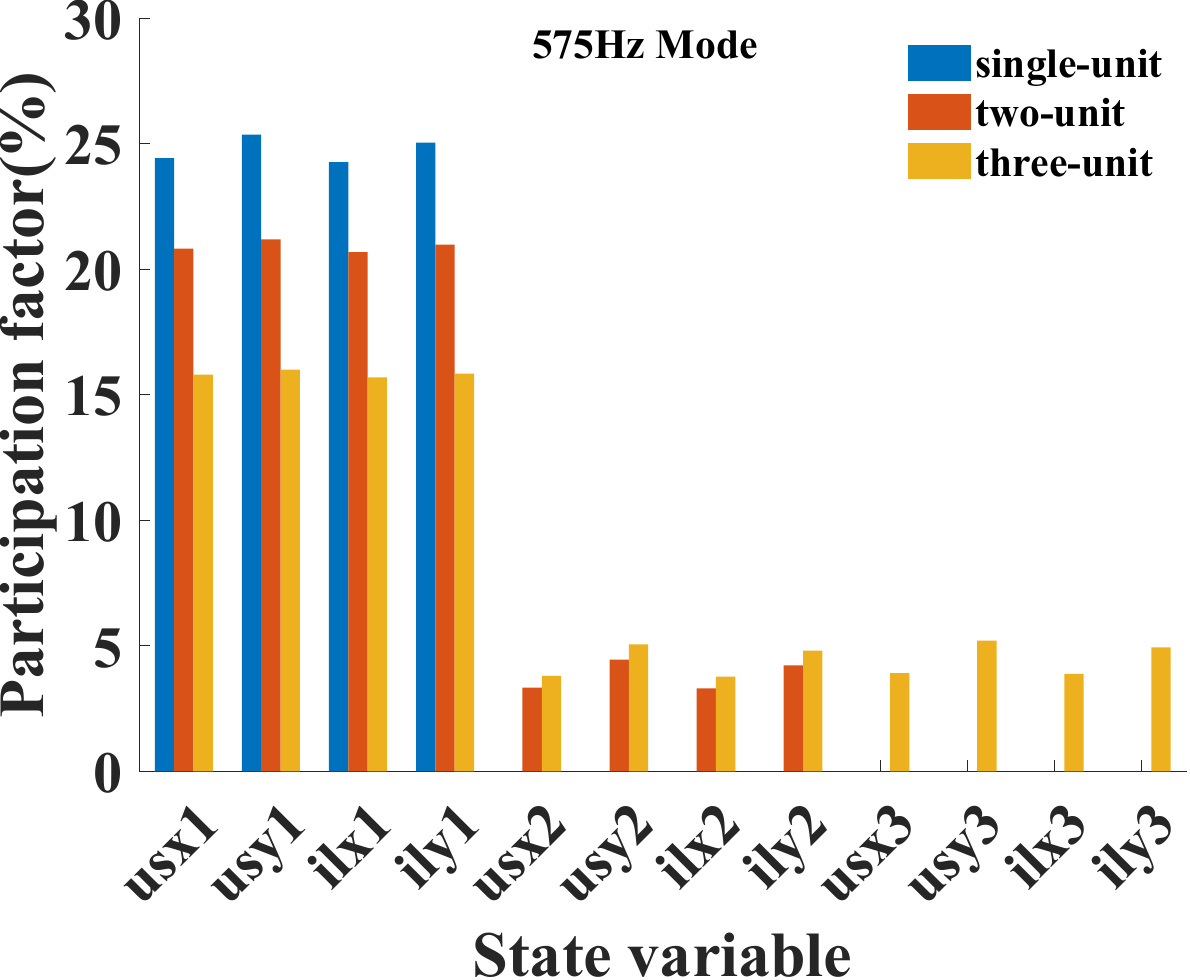}
    \caption{Participation factor}\label{fig:d}
\end{subfigure}

\caption{Critical modes switching at the stability boundary in
$k_{p41}-k_{p51}$ (normalized parameter plane), the original parameter ranges are 
$\tilde{k}_{p41} \in [0.01,100]$ and 
$\tilde{k}_{p51} \in [0.01,100]$
}
\label{fig:CriticalModesSwitching}
\end{figure}

As shown in Fig.\ref{fig:CriticalModesSwitching} (a), relative to the single-unit system, the stability regions of the two- and three-unit systems expand in the $k_{p41}-k_{p51}$ parameter plane with pronounced cusps towards different directions.
Here, $k_{p41}$ and $k_{p51}$ denote the proportional gains of unit 1's PI4 (RSC inner current loop) and PI5 (GSC outer dc-link voltage loop), respectively.
This result further reveals the nonlinear evolution of the system boundary as the number of units increases.

Similarly, for a single-machine aggregated model with more number of units, the corresponding stability boundary is shown in Fig.\ref{fig:CriticalModesSwitching} (b).
In contrast to granular models, \textit{as the number of units increases in the aggregated model, its stability boundary varies mildly, whereas the critical instability frequency decreases significantly.} 

More importantly, a switching behavior of critical modes appears on the stability boundary—not only between distinct boundary segments but also within the same segment, as shown in Fig.\ref{fig:CriticalModesSwitching} (a). Specifically, along the right-hand boundary of the single- and two-unit cases, alternating instability bands are observed at approximately 475 Hz and 575 Hz. However, as unit number increases, the 575 Hz band narrows and is absent in the three-unit system, as shown more clearly
in Fig.\ref{fig:CriticalModesSwitching} (c).
This implies that \textit{the same parameter-variation direction can drive distinct instability types in single- versus multi-unit models.} 
However, in a single-unit aggregated model (Fig.\ref{fig:CriticalModesSwitching} (b)), as the number of units increases, the unstable modal switching persists, but the oscillation frequency varies. 
Furthermore, a participation factor analysis of the 575-Hz mode, shown in Fig.\ref{fig:CriticalModesSwitching} (d), reveals that as more granular units are added to the system, state variables from these units begin to contribute. 
It suggests that \textit{as more units participate, it is  difficult for a single aggregated model, obtained by the aggregation method in this paper, to accurately replicate the system’s participation.}

\subsection{Coupling characteristics on the stability boundary}

\begin{figure}[tb]
\centering
\begin{subfigure}[t]{0.49\linewidth}
     \includegraphics[width=\linewidth]{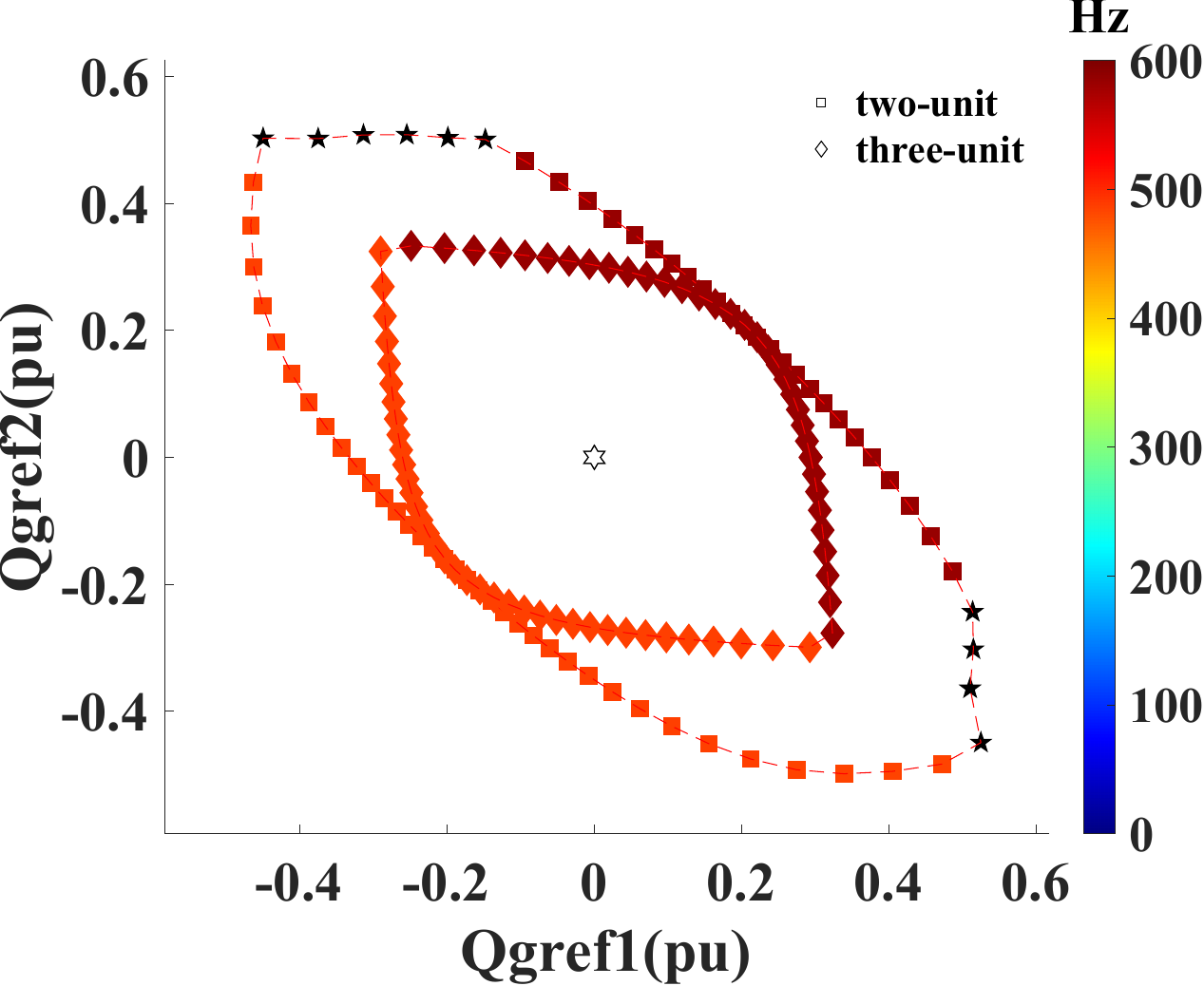}
    \caption{Stability boundary of granular models}\label{fig:a}
\end{subfigure}\hfill
\begin{subfigure}[t]{0.49\linewidth}
    \includegraphics[width=\linewidth]{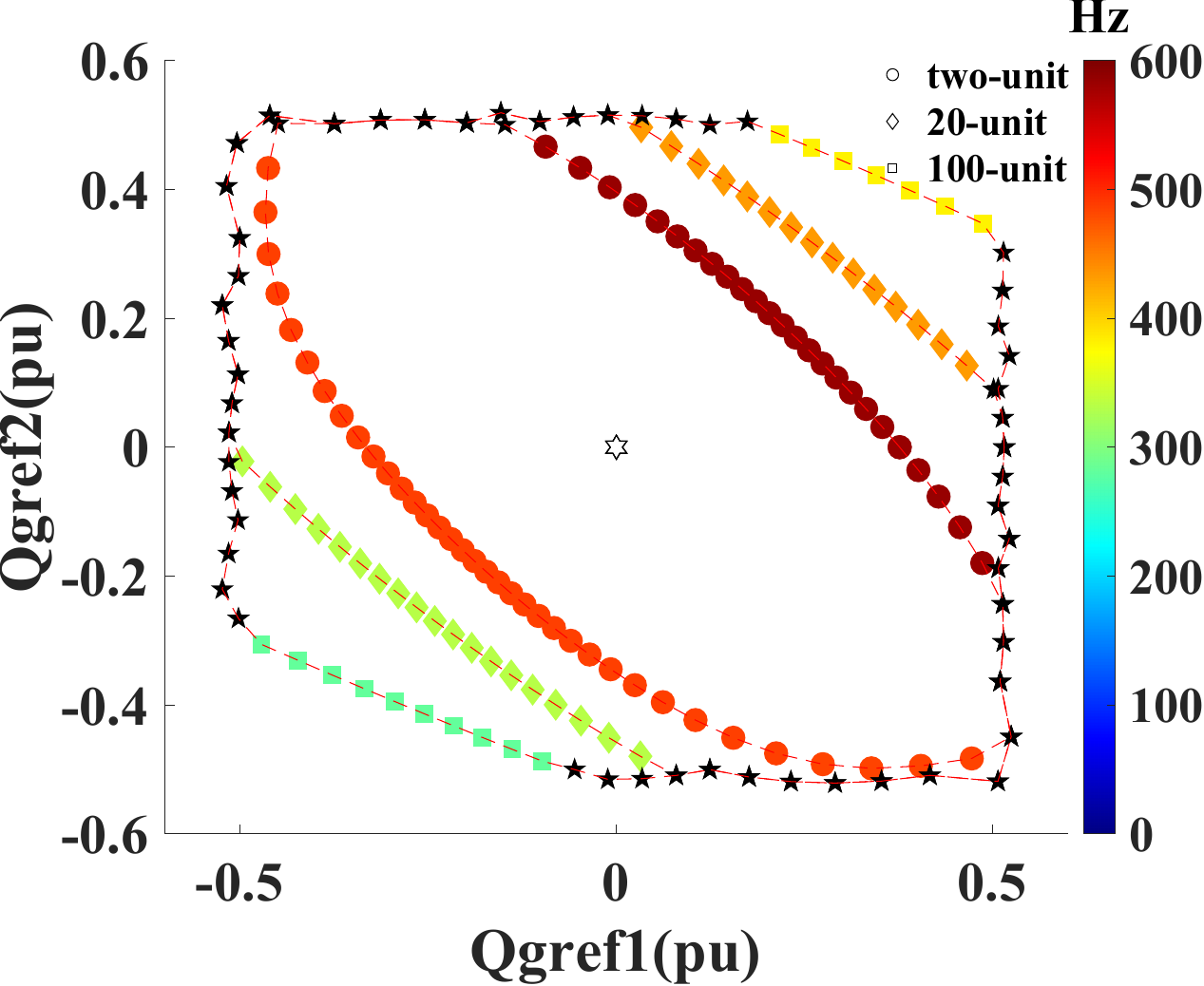}
    \caption{Stability boundary of single-unit aggregated models}\label{fig:b}
\end{subfigure}
\begin{subfigure}[t]{0.49\linewidth}
     \includegraphics[width=\linewidth]{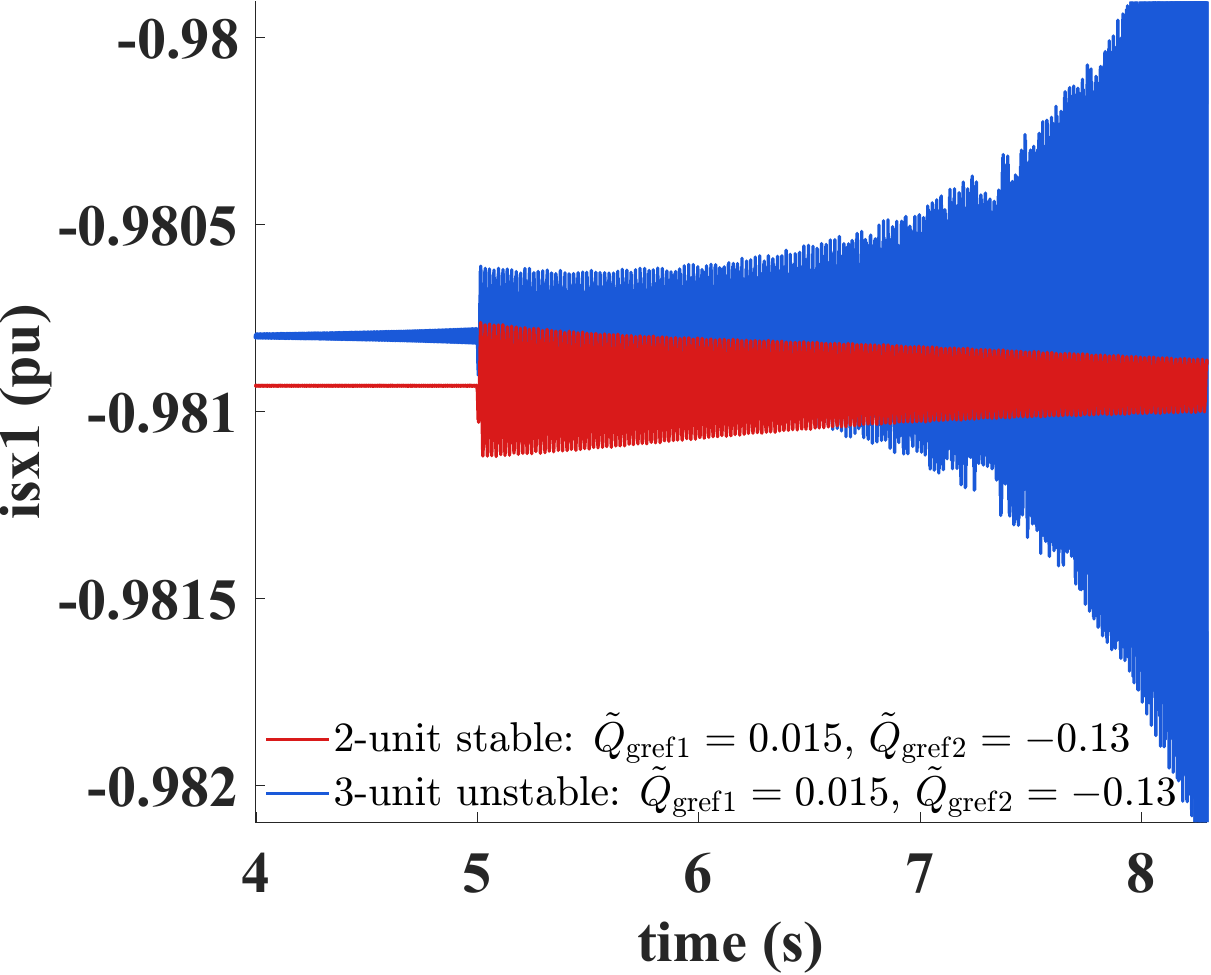}
    \caption{Time-domain response}
    \label{fig:c}
\end{subfigure}\hfill
\begin{subfigure}[t]{0.49\linewidth}
\includegraphics[width=\linewidth]{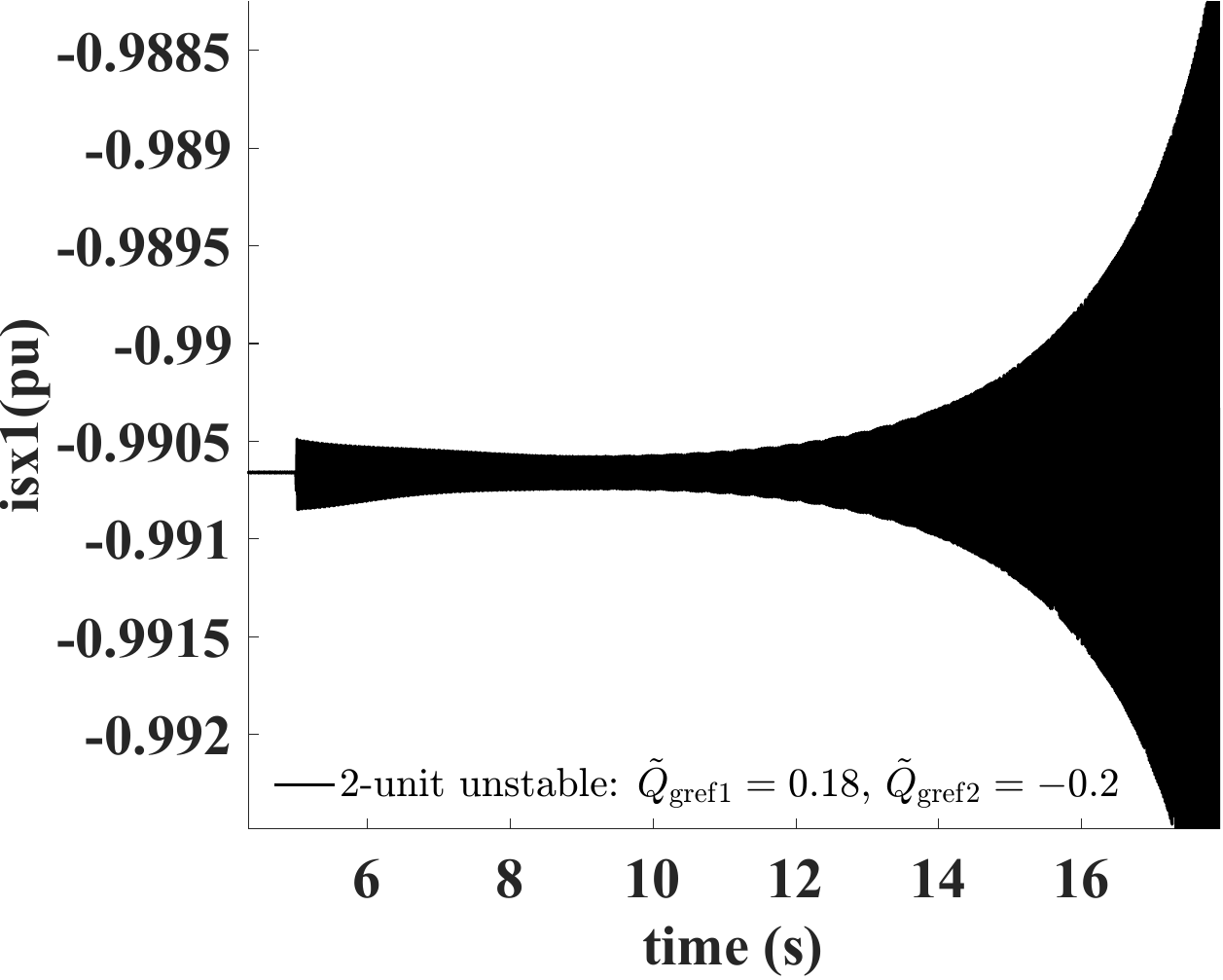}
    \caption{Time-domain response}\label{fig:d}
\end{subfigure}

\caption{Coupling characteristics at the stability boundary in
$Q_{gref1}-Q_{gref2}$ (normalized parameter plane), the original parameter ranges are 
$\tilde{Q}_{gref1}\text{(p.u.)} \in [-0.2,+0.2]$ and 
$\tilde{Q}_{gref2}\text{(p.u.)}\in [-0.2,+0.2]$
}
\label{fig:couplingcharacteristics}
\end{figure}

As shown in Fig.\ref{fig:couplingcharacteristics} (a) and (b), the stability boundary of the operating parameters exhibits significant complementary regulatory coupling characteristics.
Specifically, in the $Q_{gref1}-Q_{gref2}$ plane (representing the reactive power setpoints of the Unit 1 and Unit 2, respectively), this boundary does not show a regular distribution under independent parameter constraints, but rather exhibits significant diagonal extension and symmetrical features. In particular, the system's total reactive power reaches an extremum near the 45-degree line, while a negatively correlated stable region for reactive power allocation forms near the -45-degree line, indicating a certain complementary regulatory capacity between the two units.
Therefore, the operational stability of the system is jointly determined by the combined variations of multiple operating parameters. When one parameter shifts towards instability, another can compensate through reverse regulation within a certain range, thereby keeping the operating point within the stable region.

As shown in Fig.\ref{fig:couplingcharacteristics} (c), the two-unit and three-unit systems were tested with the parameters set to $Qgref1=0.015$ and $Qgref2=-0.13$, which corresponds to the point (0.0375,-0.325) in the normalized plane. Under these conditions, the three-unit system exhibits oscillatory instability, whereas the two-unit system remains stable. When the parameters are further adjusted to $Qgref1=0.18$ and $Qgref2=-0.2$, corresponding to (0.45,-0.5) in the normalized plane, the two-unit system also becomes unstable, as illustrated in Fig.\ref{fig:couplingcharacteristics} (d). These time-domain simulations further verify the accuracy of the parameter-sweep results.

Further comparison reveals that in the detailed model, the parameter-plane stability region of the 3-unit system exhibits more pronounced shrinking and distortion compared to the 2-unit system, resulting in a significantly reduced dispatchable stable operating range. In contrast, because the aggregated model is fundamentally based on equivalent parameter scaling, its stability region evolves only slowly from its initial shape when the number of units increases. It fails to accurately reflect the significant boundary shrinkage caused by the enhanced dynamic coupling among units observed in the detailed model.

Consequently, the aforementioned differences between the models cannot be ignored; otherwise, they may lead to a significant risk of misjudging stability margins in engineering applications.

\section{Conclusion}
\label{Conclusion}
This paper presented a thorough investigation of the small-signal stability boundaries for aggregated DFIG models and their granular counterparts. 
A unified mathematical formulation is developed to delineate stability boundaries in multidimensional parameter space. Then, a ray-based extrapolation and boundary correction algorithm is proposed to generate stability regions under parameter variations.

Two-dimensional parameter slices were visualized to reveal pronounced nonconvexity in the DFIG stability region. The boundary is not determined by a single mode: different unstable modes can switch, implying that a universal bandwidth mitigation may not be possible. 
Comparative analysis shows that 
active modes in the single-unit system may not be excited in multi-unit systems. 
Complementary regulation and coupling effects may exist among the setting parameters of multi-unit operating conditions. Consequently, the schedulable security and stability region of the system exhibits distinct anisotropic distribution characteristics.

Given the limitations of aggregated models in capturing stability boundaries, our future work will leverage the proposed characterization tool to develop an efficient computational framework for precisely quantifying the small-signal stability margins of granular wind farms.

\bibliographystyle{IEEEtran}
\bibliography{Reference}
\end{document}